\documentclass[conference]{IEEEtran}
\IEEEoverridecommandlockouts
\usepackage{cite}
\usepackage{amsmath,amssymb,amsfonts}
\usepackage{algorithmic}
\usepackage{graphicx}
\usepackage{textcomp}
\usepackage{booktabs}
\usepackage{url}
\usepackage{hyperref}
\usepackage{xcolor}
\def\BibTeX{{\rm B\kern-.05em{\sc i\kern-.025em b}\kern-.08em
    T\kern-.1667em\lower.7ex\hbox{E}\kern-.125emX}}
\begin{document}

\title{Unified View of IoT and CPS Security and Privacy\\
}


\author{
\IEEEauthorblockN{Lan Luo\IEEEauthorrefmark{1},
Christopher Morales-Gonzalez\IEEEauthorrefmark{4},
Shan Wang\IEEEauthorrefmark{2}\IEEEauthorrefmark{3},
Zhen Ling\IEEEauthorrefmark{3},
Xinwen Fu\IEEEauthorrefmark{4}
}
\IEEEauthorblockA{\IEEEauthorrefmark{1}Anhui University of Technology. Email: Email:lluo@ahut.edu.cn} 
\IEEEauthorblockA{\IEEEauthorrefmark{4}University of Massachusetts Lowell. Email: christopher\_moralesgonzalez@student.uml.edu, xinwen\_fu@uml.edu}
\IEEEauthorblockA{\IEEEauthorrefmark{2}Hong Kong Polytechnic University. Email: shan-cs.wang@polyu.edu.hk}
\IEEEauthorblockA{\IEEEauthorrefmark{3}Southeast University. Email: zhenling@seu.edu.cn}
}

\maketitle

\begin{abstract}
The concepts of Internet of Things (IoT) and Cyber Physical Systems (CPS) are closely related to each other. IoT is often used to refer to small interconnected devices like those in smart home while CPS often refers to large interconnected devices like industry machines and smart cars. In this paper, we present a unified view of IoT and CPS: from the perspective of network architecture, IoT and CPS are similar given that they are based on either the OSI model or TCP/IP model. In both IoT and CPS, networking/communication modules are attached to original things so that isolated things can be integrated into cyber space. If needed, actuators can also be integrated with a thing so as to control the thing. With this unified view, we can perform risk assessment of an IoT/CPS system from six factors, hardware, networking, operating system (OS), software, data and human. To illustrate the use of such risk analysis framework, we analyze an air quality monitoring network, smart home using smart plugs and building automation system (BAS). We also discuss challenges such as cost and secure OS in IoT security.
\end{abstract}


\section{Introduction}

While the definitions of Internet of Things (IoT) and Cyber Physical Systems (CPSs) may vary, they are converging. Generally, both systems consist of physical devices with embedded sensors and/or actuators, perform computations on the data reported and are linked through networks to automate tasks. Where they diverge is their primary application areas. IoT is often used to refer to small interconnected devices like those in smart home while CPS often refers to large interconnected devices like industry machines and smart cars.

When it comes to network communications, the underlying mechanisms of both IoT and CPS networks are quite similar. Both use the OSI and/or TCP/IP network models.
Devices in IoT networks often utilize IP addresses for communication within or between IoT networks.
CPSs often rely on wired or non-IP based protocols, often in isolated networks. Advancements have introduced IP-based communication in CPSs, exemplified by protocols like KNX/IP, which is a building automation protocol and enables the old fashioned KNX communications to be sent over an IP backbone.

\subsection{Internet of Things (IoT)}
IoT networks can have objects which can be physical or even virtual ones. An object shall be addressable. For example each object can have an IP so that we can communicate with the object. So the key point here is the interconnected objects, which can be anything. It can be a humidity and temperature sensor, a smart plug, a smart camera, and many other things. Those things generate data, which can be uploaded into the cloud for further processing. In this way, things become smarter since we can make decision based on the data.

\begin{figure}[t]
  \centering
\includegraphics[width=0.95\columnwidth]{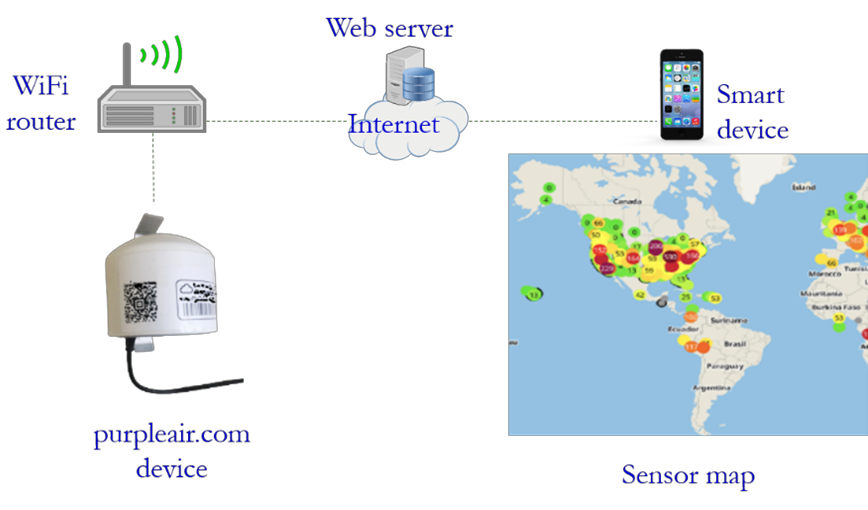}\\
    \caption{Air Quality Monitoring Network}\label{fig::fig1} 
\end{figure}

Fig. \ref{fig::fig1} shows an example IoT system. This is a low-cost air quality monitoring network \cite{luo2018security}. An air quality monitoring device is connected to a WiFi router and sends air quality data (PM2.5) to a server, which publishes the data via a web server and online map, visualizing the air quality at locations where the air quality devices are deployed. Computers and smartphones can be used to access the map. The air quality device itself contains a microcontroller (MCU)---ESP8266---and an air quality sensor. The MCU reads the air quality sensor data via UART and sends the data out via its internal WiFi chip. 
The \textit{thing} in this IoT air quality monitoring network example is the air quality sensor which is connected via UART to the MCU with WiFi. The MCU is used to read the sensor and send the data to cyber space via WiFi.


\subsection{Cyber Physical System (CPS)}
\begin{figure}[t]
  \centering
    \includegraphics[width=1\columnwidth]{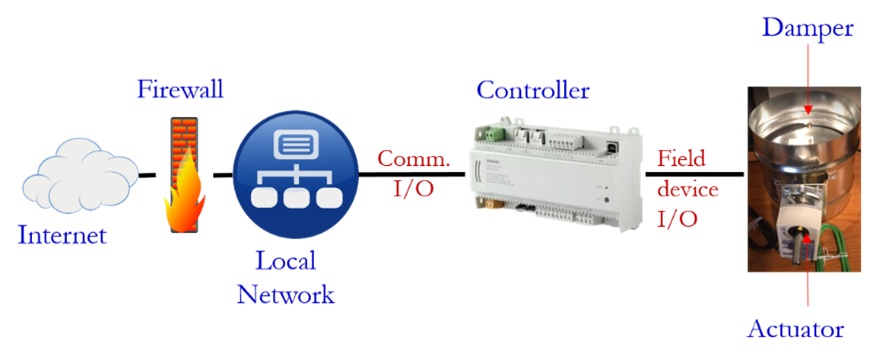}\\
    \caption{Air Quality Monitoring Network}\label{fig::fig2} 
\end{figure}

Fig. \ref{fig::fig2} shows an example cyber physical system (CPS) in the form of a  smart building. The physical thing is the damper. The angle of the damper blade can be adjusted to control the airflow. The damper has a handle, which turns and adjusts the blade. The handle can be fastened to a damper actuator, which rotates the handle. The actuator can be connected to a controller in multiple ways. For example, SIEMENS DXR2.E12P-102P is a compact room automation station (controller). A damper actuator can be connected to the controller through a proprietary TX-I/O module for power supply and communication. The actuator can also be connected to the controller through the building automation protocol KNX. Since KNX is available on the actuator, the actuator with the integrated damper can form a KNX network with other KNX-capable building components. DXR2.E12P-102P is basically a specialized computer and has networking capabilities such as TCP/IP, BACnet/IP and KNX. It can be connected to a local area network (LAN). The LAN is normally protected by a firewall and communicates with the Internet.

A demo of controlling the damper can be found at \url{https://youtu.be/KOm9banQ86g}. We attach an actuator to the damper. The actuator has I/O modules connecting to a controller, which can then control the actuator and thus the damper. In this particular case, the actuator supports the BAS protocol KNX and allows the actuator plus damper to form a KNX network with other KNX devices. The controller has the network connectivity, connecting to the Internet through local network and firewall. Please refer to \cite{humayed2017cyber} for other examples of CPS.

\section{Unified View of IoT and CPS}

\subsection{Unified Network Architecture}
We are interested in the cybersecurity of IoT and CPS and particularly focus on networking and computer systems. From the perspective of networking, IoT and CPS are similar since they are based on either the OSI model or TCP/IP model. An air quality sensor without cyber capability may show readings on a display attached to the sensor. Now when we connect a MCU with the WiFi capability to the air quality sensor (i.e., a thing) via UART, we create a smart air quality monitoring device, which now has the networking capability and becomes part of the cyber space. As for the BAS, a damper without cyber capabilities may be manually controlled. When we attach an actuator with the networking capability through its KNX module to the damper (i.e., a thing), we have a smart damper with the networking capability, which can form a BAS network through KNX. When the actuator is connected to the controller, the controller becomes a gateway to the Internet. 

In both examples, we add a networking-capable module and other necessary components to original isolated things so that these things can connect to a network and we may communicate and/or control the things remotely. The smart things can now form networks, which do not necessarily use TCP/IP and may use particular protocols such as KNX and BACnet in building automation. The network can get very complicated depending on the application. For example, in a smart building application, there are many things such as heating, ventilation, and air conditioning devices. Those things can be networked together via building automation protocols. However, if it is not a TCP/IP network, gateways are needed for the specific network to communicate with the Internet.

\begin{figure}[t]
  \centering
\includegraphics[width=1\columnwidth]{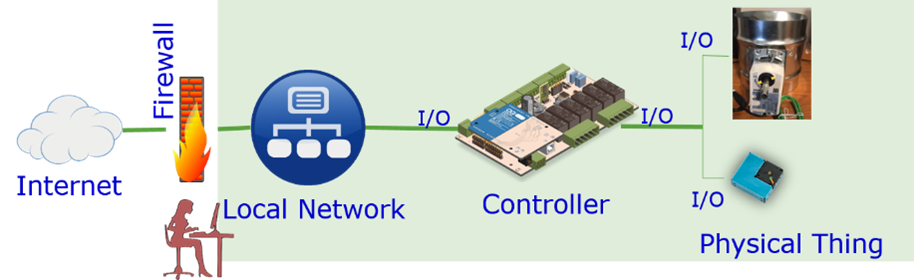}\\
    \caption{Unified Architecture of IoT and CPS}\label{fig::fig3} 
\end{figure}

Fig. \ref{fig::fig3} shows the unified view of the IoT and CPS architecture from the networking perspective. In this architecture, we have the Internet, firewall, a local area network, controller and things. Things are connected to the controller via various I/Os. An I/O may have the networking capability or not. For example, UART does not have the networking capability while KNX has the networking capability. If things use network I/O, these things can form a network on their own. Otherwise, things can form a network through the controller, which has the networking capability in general.

\subsection{Difference between IoT and CPS}
The difference between CPS and IoT is the applications. When we talk about CPS, we often talk about industrial control systems (ICS), smart grid systems, smart cars and smart buildings. The physical things in those applications are often large. When we talk about IoT, we talk about smart plugs, smart bulbs and smart cameras. The physical things in such applications are usually small. Because of the different physical things, the communication modules and actuators added to the physical things can be different. Different applications may use different hardware, different OSs, software and programming languages. The networking may use specific protocol other than the common TCP/IP. An ICS may use the Modbus or Distributed Network Protocol (DNP3) protocol. A smart grid may use the Modbus, DNP3 or Inter-Control Center Protocol (ICCP). Medical devices may use wireless protocol Bluetooth or ZigBee for convenience of surgical procedures. Smart cars may use the Controller Area Network (CAN)
and Local Interconnect Network (LIN) to interconnect different components in the car.

\section{Risk Analysis Based on Unified View}
\begin{figure}[t]
  \centering
    \includegraphics[width=0.6\columnwidth]{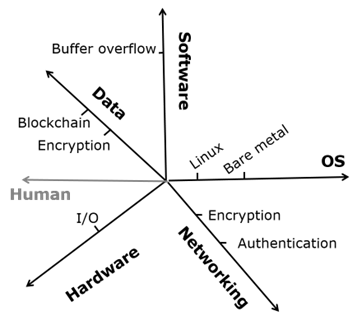}\\
    \caption{Risk analysis from six factors}\label{fig::fig4} 
\end{figure}

When we perform risk assessment of an IoT or CPS system, we can investigate six factors as shown in Fig. \ref{fig::fig4}: hardware, networking, operating system (OS), software, data and human. 

The \textit{hardware} is the physical thing. Attackers may attack the hardware I/O of a thing. For example, an appropriate cable can be attached to the hardware for the purpose of accessing the firmware of the thing and hacking the system. For \textit{networking}, is encryption used for communication? Is there any authentication? Is the protocol vulnerable? The \textit{OS} is a very interesting factor. For many those resource constrained devices, they may or may not run a real OS like embedded Linux. A software development kit (SDK) may be provided by a hardware vendor for programming. Then does the OS or SDK have vulnerabilities? The \textit{software} is the user application, which may use a SDK for coding. What programming language is used for coding? Does the software have issues such as buffer overflow? For \textit{data}, is flash encryption used to protect the data on storage media? \textit{Humans} are often the weakest link in a system. 

We now use the six factors and present a few attacks from the perspective of unified view including attacks against real-world air quality monitoring devices, smart plugs and smart building.

\subsection{Attacks against IoT}
\begin{figure}[t]
  \centering
\includegraphics[width=0.6\columnwidth]{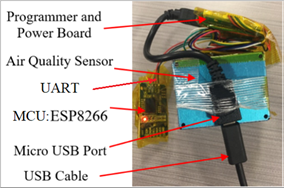}\\
    \caption{Inside of the air quality monitoring device}\label{fig::fig5} 
\end{figure}

We use attacks against air quality monitoring devices to demonstrate the attacks against IoT.
Fig. \ref{fig::fig5} shows the inside of the air quality monitoring device.

\textbf{Hardware attack.} 
When we attach a common smartphone micro USB cable which supports power supply and communication to the device, we can dump the flash content including firmware out of this device, which uses the microcontroller ESP8266 from Espressif Systems \cite{gao2019microcontroller}. ESP8266 does not have any kind of hardware protection.
The flash content includes the WiFi credential.

\textbf{Networking attack---data pollution attack.}
To understand the protocol, we construct a testbed \cite{luo2018security}. A laptop is set up as a WiFi router and installed with a protocol analysis tool called the \textit{mitmproxy}, which can capture the passing data from the sensor to its server. Because the data is not encrypted, we can analyze all the data and find the air quality monitoring system architecture, communication protocol, and data format. We find that the MAC address is used as the identity of the device and there is no real authentication or other security measures. Once we understand the protocol, we can perform the data pollution attack, creating a fake software device and injecting fake data from the fake device into the victim server.  
Our experiments show that we can change the air particle PM2.5 reading dramatically.

\subsection{Attacks against CPS}
We use the six factors in Fig. \ref{fig::fig4} to perform risk analysis of IoT systems above. We can also use the six factors to perform risk assessment of a CPS system while we think it is not that necessary to differentiate IoT and CPS. 

\begin{figure}[t]
  \centering
    \includegraphics[width=1\columnwidth]{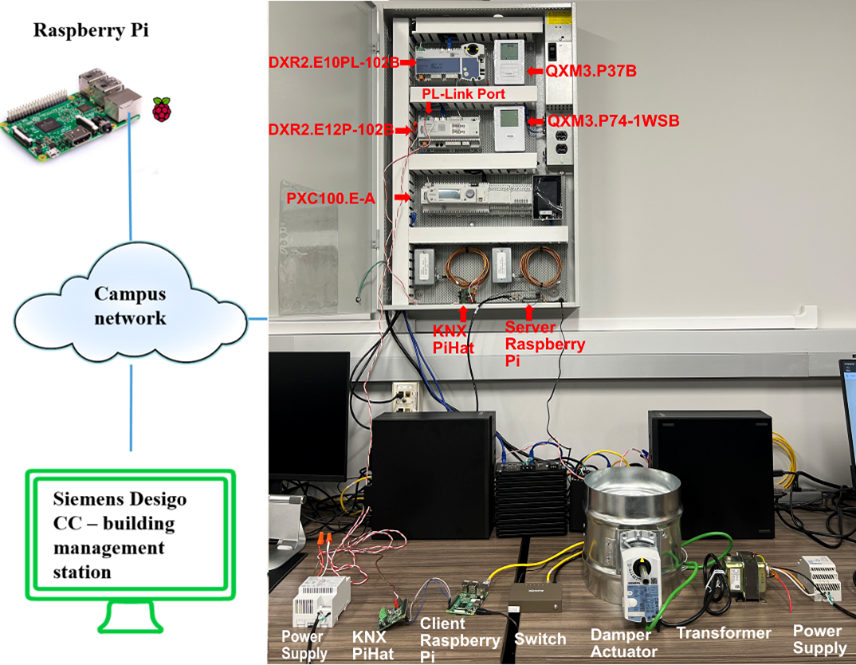}\\
    \caption{Smart Building Testbed}\label{fig::fig8} 
\end{figure}

Fig. \ref{fig::fig8} shows a smart building system panel installed by SIEMENS \cite{cash2023false}. The two QXM3 devices with the screens are temperature and humidity sensors. The two DXR2 devices on the top right are the building automation controllers, which collect readings from these two temperature sensors and send them to the Desigo CC server, a building management station software program. The Desigo CC server can visualize the layout of BAS components and perform operations if needed. The Desigo CC uses the BAS protocol BACnet/IP to communicate with the controllers. We reverse engineer the panel and are able to know the protocol details. With such information, we can use the Raspberry Pi on the top left, instead of the Desigo CC, to communicate with the controllers too. This building automation system doesn't have much security. 

We purchased a lot of other building automation system components and attached them to the panel. The bottom of Fig. \ref{fig::fig8} shows a damper with an attached actuator, KNX Pi Hats, KNX power supplies and a power transformer for the actuator. The KNX power supply provides DC power while the actuator requires AC power. Therefore, the KNX power supply is connected to the transformer, which is then connected to the actuator.
\looseness=-1

\begin{figure}[t]
  \centering
    \includegraphics[width=1\columnwidth]{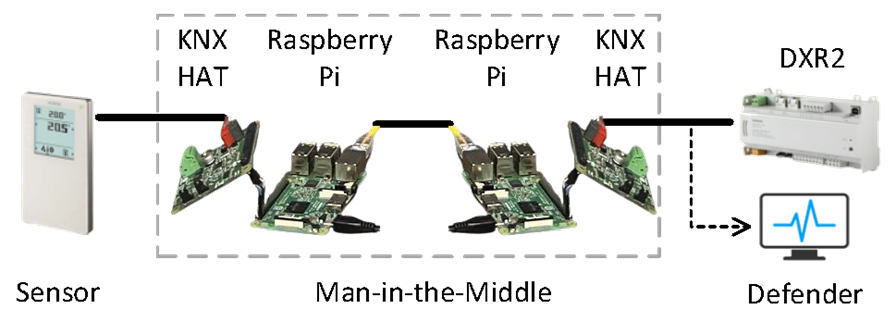}\\
    \caption{Man in the middle attack against BAS}\label{fig::fig10} 
\end{figure}

\textbf{Hardware attack against BAS.} In this example attack, we target the KNX temperate sensor on the panel. We detach the wire of the temperature sensor from the panel. We then connect the detached wire to our Raspberry Pi so as to inject data or commands into the BAS system. Such an attack is realistic since sensors like the temperature sensor are often deployed all over a building. How can we actually hook a Pi to the physical BAS? We have to understand the physical connection. Our temperature sensor and BAS in this example use the BAS protocol---KNX---to communicate with each other. As shown in Fig. \ref{fig::fig10}, we use a particular adapter called KNX Pi head, which is analogous to a network card, but for KNX. The cables detached from the temperature sensor are connected to cable jacks on the KNX Pi Hat. Actually, twisted pair cables are used to connect the Hat to the BAS and Dupont wires are used to connect the Hat to the Pi. 

To understand the details of the KNX protocol, we developed a KNX bus dump software tool, which uses the combination of the Pi and KNX Pi Hat in Fig. \ref{fig::fig10} to dump raw data from twist pair cables.  Twisted pair cables that are chained together can work as a bus connecting many KNX devices together. We format the dumped data in such a way that we can import the dump file into Wireshark, which has a KNX plugin to dissect a telegram, which is the KNX’s protocol data unit analogous to a packet in the TCP/IP protocol. Wireshark cannot be used to dump KNX telegrams since Wireshark works with the TCP/IP network and KNX is not a TCP/IP protocol. The KNX bus dump tool allows us to look at the details of telegrams and understand the actual protocol and data fields. We presented this tool at Blackhat Asia 2022.


\textbf{False data injection attack.} After we understand the building automation system, the protocol and data fields of KNX telegrams, we show we can do some damage. We want to tell the building automation industry to secure their devices such as temperature, humidity and motion sensors, which may be exposed to the public and subject to abusers. For example, somebody may remove those devices, hook their own Pi or computers to the BAS, and manipulate the BAS. 

With the combination of Pi and KNX Pi Hat, we can deploy a man-in-the-middle attack as shown in Fig. \ref{fig::fig10} so as to perform the false data injection attack. One set of Pi and KNX Pi Hat is connected to the temperature sensor and a second set is connected to the DXR2 controller. The two Pis grab KNX telegrams generated at their side, and forward them to the other side whenever needed through Ethernet. Since we already understand the protocol and the data fields of the communication between the sensor and the controller, we can actually change the temperature values in telegrams.

\begin{figure}[t]
  \centering
    \includegraphics[width=0.7\columnwidth]{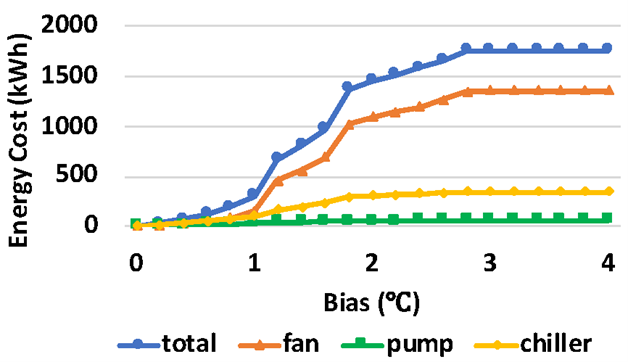}\\
    \caption{Impact of False Data Injection Attacks}\label{fig::fig11} 
\end{figure}

Fig. \ref{fig::fig11} shows the impact of the false data injection attack in terms of energy cost through simulations. Cooling is simulated here in Florida, USA. We add a small value, denoted as \textit{bias}, to the actual temperature sensor reading. Fig. \ref{fig::fig11} shows the extra energy consumption versus the bias. It can be observed that the total energy waste caused by the false data injection attack is significant. \looseness=-1

\textbf{Other attacks.} We now show a few other attacks against BAS. We just talked about attacks against KNX, which is popular in Europe. BACnet/IP is a popular BAS protocol in North America. Many buildings use insecure BACnet/IP with not much security. We designed an automatic tool to scan an IP address and determine if a BACnet device runs at that IP address. To implement such a tool, we need to understand the details of the BACnet/IP protocol. Based on the features of the BACnet/IP, we send the requests to the IP. Based on the response, we may decide if there is a BACnet/IP device over there. If yes, we can query the device for BACnet objects associated with the BACnet device. For example, if the BACnet/IP device is a controller, many other BAS devices may be connected to this controller. We can then obtain detailed information of those BAS devices and control them. For example, the SIEMENS DRX2.E10PL is a BACnet/IP device and has an integrated actuator. We are able to rotate the actuator as shown at \url{https://youtu.be/YUfO8GQILxQ} by sending commands to the DRX2.E10PL. Please note that, the actuator of the DRX2.E10PL is not connected to a damper. In this example, we just want to show security is critical to BAS. If there is no security, once people understand the protocols, they may control the BAS devices. If the devices are part of a power plant, the consequence could be disastrous. We really cannot rely on security-by-obscurity (i.e., hoping hackers do not know the protocol and network setup) for security.\looseness=-1

Software security in a BAS is also critical. BAS devices are controlled by code and software. We want to understand if the software of the building automation system devices is vulnerable to common attacks, such as buffer overflow and denial of service (DoS) attacks. We performed fuzzing against popular open source software programs including \textit{knxd}, \textit{Calimero} and \textit{Bacnet-stack} and found all of them have software bugs. In fuzzing, we send delicate junk messages into the target software and see how it behaves \cite{pearson2022fume}. If the device or the software crashes, then there is a problem. 
For example, it could be a buffer overflow vulnerability that caused the crash.
Further investigation of the code can help identify such a problem.

\section{Challenges in IoT Security and Privacy}

There are two main fields of research in IoT, attack and defense. Both attack and defense research focuses on the six factors in Fig. \ref{fig::fig4}.

\subsection{Causes of security issues in IoT}
One question we want to ask is what is the root cause of so many IoT security and privacy issues? Of course, any ignorance and incapability of the six factors can cause problems.
\looseness=-1

\textbf{IoT hardware.} It appears hardware may not be the cause of IoT security and privacy any more while many people think so. People may think we do not have appropriate secure chips for IoT and those chips are too expensive. Those are misconceptions now. There are various low-cost chips designed for securing IoT. For example, ESP32 microcontrollers from Espressif Systems support Wi-Fi,  Bluetooth, BLE, hardware crypto acceleration, flash encryption and secure boot, and are around a couple of US dollars. The ATECC608A from Microchip Technology Inc. is a crypto co-processor, which accepts data and performs cryptographic operation on the data. It is around 50 US cents and supports AES, ECC, HMAC, SHA-256, RNG (random number generator), and secure key storage.
\looseness=-1

We have tested the performance of some of those chips and they are actually pretty good. The ECC key generation on ESP32 needs about 0.23 second, which is appropriate for various low cost applications like an air quality monitoring sensors. Hard real time may not be needed for many IoT applications. ECC signature generation and verification is within 0.5 second. The public cryptography algorithms are often used for key exchange, after which we often use the symmetrical key cryptography algorithms such as AES for data encryption and decryption. AES key generation is on a scale of hundreds of microseconds. AES encryption and decryption is on a scale of microseconds. 

We believe design and ignorance are major causes of poor IoT security and privacy while hardware may not be the major issue for emerging IoT devices. Many IoT devices are pushed to the market without much consideration of security and privacy design. Training is an issue too since secure coding is not taught widely in school.

\textbf{Cost.} For large-scale applications like smart buildings and smart grids, there are so many components. It incurs huge costs on labor and new devices when we try to innovate them for the purpose of security. Therefore, one challenge in those applications is how to mitigate the danger if the physical things cannot be upgraded or can only be partially innovated. 

\subsection{Secure Operating System for IoT}
We have been actually looking at how to secure IoT applications at the level of operating system. IoT application is often written in C and C++ and such an application is vulnerable to memory corruption attacks, e.g., buffer overflow attacks. Can we use traditional address space layout randomization (ASLR) to fight memory corruption attacks in resource constrained IoT devices? One challenge is the limited SRAM and flash in resource constrained IoT devices. For example, Microchip’s SAML11 microcontroller uses ARM Cortex-M23 CPU, which runs at up to 48 MHz and has only 64KB Flash and 16KB SRAM. In Windows, Linux and MacOS, the whole executable is loaded to a random address in SRAM with ASLR. There may be no so much RAM in resource constrained IoT devices. We also do not want to sacrifice much performance given the limited computing power of those IoT devices. 

We have developed a function based ASLR for resource-constrained IoT systems \cite{luo2022faslr}. The app on the flash is marked as non-executable through the memory protection unit (MPU). Now every time a function is called, a MPU hardware exception will be raised and our particular exception handler will randomly load the function into the SRAM for execution. Such function randomization raises the issues of memory management and addressing. We have addressed those challenges and developed a real-world device with decent performance.

Can we develop operating systems for resource constrained IoT devices incorporating various security measures such as ASLR and control flow integrity (CFI)? There are various efforts on secure operating systems \cite{luo2022security}.

\section*{Acknowledgment}
This research was supported in part by US National Science Foundation (NSF) awards 1931871 and 1915780, and US Department of Energy (DOE) Award DE-EE0009152. Any opinions, findings, conclusions, and recommendations in this paper are those of the authors and do not necessarily reflect the views of the funding agencies.


\bibliographystyle{IEEEtran}

\end{document}